\documentclass[12pt]{amsart}
\usepackage{amssymb}

\theoremstyle{plain}

\numberwithin{equation}{section}

\begin{document}
\title{Supersymmetry of the Morse oscillator}
\author{Hashim A Yamani${}^1$ and Zouha\"{i}r Mouayn${}^2$}
\address{\small {${}^{1}$ King Abdullah City for Atomic and Renewable Energy, PO Box 2022, Riyadh, 11451, Saudi Arabia }}
\email{hashim.haydara@gmail.com}%
\address{\small{${}^{2}$ Department of Mathematics, Faculty of
Sciences and Technics (M'Ghila) PO Box 523, B\'{e}ni Mellal, Morocco}}
\email{mouayn@gmail.com}%
\maketitle

\begin{abstract}
While dealing in $\left[ 1\right] $ with the supersymmetry of a tridiagonal
Hamiltonian $H$, we have proved that its partner Hamiltonian $H^{\left(
+\right) }$ also have a tridiagonal matrix representation in the same basis
and that the polynomials associated with the eigenstates expansion of $%
H^{\left( +\right) }$ are precisely the kernel polynomials of those
associated with $H$. This formalism is here applied to the case of the Morse
oscillator which may have a finite discrete energy spectrum in addition to a
continuous one. This completes the treatment of tridiagonal Hamiltonians
with pure continuous energy spectrum, a pure discrete one, or a spectrum of
mixed discrete and continous parts.
\end{abstract}

\section{Introduction}

In a previous work $\left[ 1\right] $, we investigated the supersymmetry
(SUSY) of a given positive semi-definite Hamiltonian, $H$, whose matrix
representation in a chosen basis is tridiagonal. Precisely, we have defined
a forward-shift operator $A$ and its adjoint, the backward-shift operator $%
A^{\dag },$ by specifying how they act on each vector basis. \ It turns out
that these operators play a central role in our treatment. Their matrix
representations have been derived by demanding that the given Hamiltonian
has the form $A^{\dag }A$. We proved that $A^{\dag }A$ also has a
tridiagonal matrix representation in the chosen basis and we establish
explicit formulae connecting the parameters defining $A$ and the matrix
elements of $H$. These parameters are closely related to the set of
polynomials associated with the eigenstates expansion of $H$ in the chosen
basis. Writing the partner Hamiltonian $H^{\left( +\right) }$ as $AA^{\dag }$%
, we showed that it too has a tridiagonal matrix representation in the same
basis. We then have established that the set of polynomials associated with $%
H^{\left( +\right) }$ are precisely the kernel polynomials of those
associated with $H$. The applications of these results have been illustrated
to two well-know Hamiltonians, namely, the free particle Hamiltonian whose
energy spectrum is purely continuous and the harmonic oscillator Hamiltonian
whose energy spectrum is purely discrete. We confirmed that our treatment
reproduced previously established results for these two systems.

Many physical systems have energy spectra which are partly discrete and
partly continuous. In this paper we work out in some details how our
formalism is able to treat a physical system with mixed spectrum. For that
purpose we choose the case of the Morse oscillator. This system is ideal not
only because it may have a finite discrete energy spectrum in addition to
the continuous part but there exists a known basis set that renders the
Hamiltonian tridiagonal.

The Morse potential is usually written as $V\left( x\right) =V_{0}\left(
e^{-2\alpha x}-2e^{-\alpha x}\right) $ where $V_{0}$ and $\alpha $ are given
parameters that related to the properties of the physical system that this
potential is attempting to model. In particular, $V_{0}$ controls the depth
of the potential while $\alpha $ controls the its width. For the purpose of
this paper, we assume that the values of $V_{0}$ and $\alpha $ are such that
the potential possesses one or more bound states.

The paper is organized as follows. In section 2, we review briefly the
formalism developed to treat the supersymmetry of tridiagonal Hamiltonians.
In section 3, we apply this formalism to the specific case of the Morse
oscillator\textbf{.} We then identify the basis set that renders the
Hamiltonian tridiagonal. In the process of finding the representation of the
energy eigenvectors in the same basis, we solve the resulting three term
recursion relation satisfied by a set of orthogonal polynomials. The
solution includes the identification of the discrete spectrum and the form
of the completeness relation satisfied by the orthogonal polynomials. We
then find the matrix representation of the supersymmetric partner
Hamiltonian and show, in an analogous manner, how to completely characterize
its properties. \ In section 4, we discuss the obtained results and derive
additional known ones to suggest that our formalism is a viable tool in the
study of supersymmetry.\ \ \ \ \ \ \ \ \ \ \ \ \ \ \ \

\section{Summary of results on supersymmetry of tridiagonal Hamiltonians}

In this section we summarize the results of our previous work on
supersymmetry of tridiagonal Hamiltonians $\left[ 1\right] $.

We assume that the matrix representation of the given Hamiltonian $H$ in a
complete orthonormal basis $\mid \phi _{n}>,n=0,1,2,...$ $,$ is tridiagonal.
That is
\begin{equation}
\left\langle \phi _{n}\mid H\mid \phi _{m}\right\rangle =b_{n-1}\delta
_{n,m+1}+a_{n}\delta _{n,m}+b_{n}\delta _{n,m-1}\text{.}  \tag{$2.1$}
\end{equation}
We now define the forward-shift operator $A$ by its action on the basis $%
\mid \phi _{n}>$ as follows
\begin{equation}
A\mid \phi _{n}>=c_{n}\mid \phi _{n}>+d_{n}\mid \phi _{n-1}>  \tag{2.2}
\end{equation}
for every $n=1,2,...$ . For $n=0$, we state that $d_{0}=0.$ Furthermore, we
require from the adjoint operator $A^{\dagger }$ to act on the ket vectors $%
\mid \phi _{n}>$ in the following way:
\begin{equation}
A^{\dagger }\mid \phi _{n}>=c_{n}\mid \phi _{n}>+d_{n+1}\mid \phi _{n+1}>%
\text{, \ \ \ \ \ \ \ \ \ }n=0,1,2,3,...\text{ .}  \tag{2.3}
\end{equation}
The operator $A^{\dagger }A$\ now admits the tridiagonal representation
\begin{equation}
\left\langle \phi _{n}\mid A^{\dagger }A\mid \phi _{m}\right\rangle
=c_{m}d_{m+1}\delta _{n,m+1}+\left( c_{m}c_{m}+d_{m}d_{m}\right) \delta
_{n,m}+d_{m}c_{m-1}\delta _{n,m-1}  \tag{2.4}
\end{equation}
in terms of the coefficients $\left( c_{n},d_{n+1}\right) ,n=0,1,2,...$\ .

We have proved that the coefficients\ in (2.1) are connected to those in $%
\left( 2.2\right) $ by the relations
\begin{equation}
a_{n}=c_{n}c_{n}+d_{n}d_{n}  \tag{2.5}
\end{equation}
\begin{equation}
b_{n}=c_{n}d_{n+1}  \tag{2.6}
\end{equation}
for every integer $n=0,1,2,...$ .

We have shown that the partner Hamiltonian $H^{\left( +\right) }=AA^{\dagger
}$\ has a tridiagonal representation with respect to the basis $\mid \phi
_{n}>,$ with the form
\begin{equation}
\left\langle \phi _{n}\mid H^{\left( +\right) }\mid \phi _{m}\right\rangle
=b_{n-1}^{\left( +\right) }\delta _{n,m+1}+a_{n}^{\left( +\right) }\delta
_{n,m}+b_{n}^{\left( +\right) }\delta _{n,m-1}  \tag{2.7}
\end{equation}
with the coefficients having the explicit values
\begin{equation}
a_{n}^{\left( +\right) }=c_{n}c_{n}+d_{n+1}d_{n+1},  \tag{2.8}
\end{equation}
\begin{equation}
b_{n}^{\left( +\right) }=c_{n+1}d_{n+1}.  \tag{2.9}
\end{equation}
The tridiagonal matrix representation of the Hamiltonian $H$ with respect to
the basis $\left| \phi _{n}\right\rangle $ also means that $H$ acts on the
elements of this basis as
\begin{equation}
H\left| \phi _{n}\right\rangle =b_{n-1}\left| \phi _{n-1}\right\rangle
+a_{n}\left| \phi _{n}\right\rangle +b_{n}\left| \phi _{n+1}\right\rangle ,
\tag{2.10}
\end{equation}
$n=0,1,2,...$ . We may then considered the solutions of the eigenvalue
problem
\begin{equation}
H\left| \varphi _{E}\right\rangle =E\left| \varphi _{E}\right\rangle
\tag{2.11}
\end{equation}
by expanding the eigenvector $\left| \varphi _{E}\right\rangle $ in the
basis $\left| \phi _{n}\right\rangle $ with real coefficients as
\begin{equation}
\left| \varphi _{E}\right\rangle =\sum\limits_{n=0}^{+\infty }\mathcal{C}%
_{n}\left( E\right) \left| \phi _{n}\right\rangle .  \tag{2.12}
\end{equation}
Then, making use of $\left( 2.10\right) $, one readily obtains the following
recurrence representation of the expansion coefficients
\begin{equation}
E\mathcal{C}_{0}\left( E\right) =a_{0}\mathcal{C}_{0}\left( E\right) +b_{0}%
\mathcal{C}_{1}\left( E\right) ,  \tag{2.13}
\end{equation}
\begin{equation}
E\mathcal{C}_{n}\left( E\right) =b_{n-1}\mathcal{C}_{n-1}\left( E\right)
+a_{n}\mathcal{C}_{n}\left( E\right) +b_{n}\mathcal{C}_{n+1}\left( E\right) ,
\tag{2.14}
\end{equation}
for every integer $n=1,2,...$ , and the orthogonality relation
\begin{equation}
\delta _{n,m}=\sum_{\mu }\mathcal{C}_{n}\left( E_{\mu }\right) \mathcal{C}%
_{m}\left( E_{\mu }\right) +\int\limits_{\Omega _{c}\left( E\right) }%
\mathcal{C}_{n}\left( E\right) \mathcal{C}_{m}\left( E\right) dE\text{.}
\tag{2.15}
\end{equation}
This relation admits the possibility that in the case when the spectrum of
the Hamiltonian $H$ is composed by a discrete part $\left\{ E_{\mu }\right\}
_{\mu }$ and a continuous part $\Omega _{c}\left( E\right) $. Define
\begin{equation}
\mathcal{P}_{n}\left( E\right) =\frac{\mathcal{C}_{n}\left( E\right) }{%
\mathcal{C}_{0}\left( E\right) }\text{, \ }n=0,1,2,...\text{ .}  \tag{2.16}
\end{equation}
Then $\left\{ \mathcal{P}_{n}\left( E\right) \right\} $ is a set of
polynomials that satisfy the three-term recursion relation for $n\geq 1$
\begin{equation}
E\mathcal{P}_{n}\left( E\right) =b_{n-1}\mathcal{P}_{n-1}\left( E\right)
+a_{n}\mathcal{P}_{n}\left( E\right) +b_{n}\mathcal{P}_{n+1}\left( E\right)
\tag{2.17}
\end{equation}
with initial conditions $\mathcal{P}_{0}\left( E\right) =1$ and $\mathcal{P}%
_{1}\left( E\right) =\left( E-a_{0}\right) b_{0}^{-1}.$

Now, with the help of the above notations, the coefficients $\left(
c_{n},d_{n}\right) $\ can also be expressed in terms of coefficients $b_{n}$%
\ and the values at zero of consecutive polynomials $\mathcal{P}_{n}$ for $%
n\geq 0$ as\textit{\ }
\begin{equation}
\left( d_{n+1}\right) ^{2}=-b_{n}\frac{\mathcal{P}_{n}\left( 0\right) }{%
\mathcal{P}_{n+1}\left( 0\right) }\text{ \ \ \ \ \ \ \ \ \ }  \tag{2.18}
\end{equation}
and
\begin{equation}
\left( c_{n}\right) ^{2}=-b_{n}\frac{\mathcal{P}_{n+1}\left( 0\right) }{%
\mathcal{P}_{n}\left( 0\right) }\text{.}  \tag{2.19}
\end{equation}
Since the partner Hamiltonian $H^{\left( +\right) }$ has also a tridiagonal
representation in the same basis, an analogous treatment yields the
following three-term recursion relation for $n\geq 1$
\begin{equation}
E\mathcal{P}_{n}^{\left( +\right) }\left( E\right) =b_{n-1}^{\left( +\right)
}\mathcal{P}_{n-1}^{\left( +\right) }\left( E\right) +a_{n}^{\left( +\right)
}\mathcal{P}_{n}^{\left( +\right) }\left( E\right) +b_{n}^{\left( +\right) }%
\mathcal{P}_{n+1}^{\left( +\right) }\left( E\right)  \tag{2.20}
\end{equation}
with initial conditions $\mathcal{P}_{0}^{\left( +\right) }\left( E\right)
=1 $ and $\mathcal{P}_{1}^{\left( +\right) }\left( E\right) =\left(
E-a_{0}^{\left( +\right) }\right) \left( b_{0}^{\left( +\right) }\right)
^{-1}$.

Here, we have assumed that coefficients $\mathcal{C}_{n}^{\left( +\right)
}\left( E\right) $ of the expansion of the eigenvector $\left| \varphi
_{E}\right\rangle ^{\left( +\right) }$ of $H^{\left( +\right) }$ , which
satisfy analog orthogonal relations to those in $\left( 2.15\right) $ as
\begin{equation}
\delta _{n,m}=\sum_{\mu }\mathcal{C}_{n}^{\left( +\right) }\left( E_{\mu
}\right) \mathcal{C}_{m}^{\left( +\right) }\left( E_{\mu }\right)
+\int\limits_{\Omega _{c}^{\left( +\right) }\left( E\right) }\mathcal{C}%
_{n}^{\left( +\right) }\left( E\right) \mathcal{C}_{m}^{\left( +\right)
}\left( E\right) dE  \tag{2.21}
\end{equation}
can be written in terms of polynomials in $\left( 2.20\right) $ as
\begin{equation}
\mathcal{P}_{n}^{\left( +\right) }\left( E\right) =\frac{\mathcal{C}%
_{n}^{\left( +\right) }\left( E\right) }{\mathcal{C}_{0}^{\left( +\right)
}\left( E\right) }\text{.}  \tag{2.22}
\end{equation}
\ Finally, the polynomials $\mathcal{P}_{n}\left( E\right) $\ are connected
to their supersymmetric partners $P_{n}^{\left( +\right) }\left( E\right) $\
as follows
\begin{equation}
\mathcal{P}_{n}^{\left( +\right) }\left( E\right) =\sqrt{\frac{b_{0}\mathcal{%
P}_{1}\left( 0\right) }{b_{n}\mathcal{P}_{n}\left( 0\right) \mathcal{P}%
_{n+1}\left( 0\right) }}\mathcal{K}_{n}\left( E,0\right)  \tag{2.23}
\end{equation}
where
\begin{equation}
\mathcal{K}_{n}\left( E,0\right) =\sum\limits_{j=0}^{n}\mathcal{P}_{j}\left(
E\right) \mathcal{P}_{j}\left( 0\right)  \tag{2.24}
\end{equation}
denotes the kernel polynomial, see $\left( \left[ 2\right] \text{, p.38}%
\right) $ or $\left( \left[ 3\right] \text{, p.35}\right) $.

\section{Tridiagonal representation of the Morse oscillator}

We are now ready to apply this formalism to the case of the Morse oscillator
whose Hamiltonian is given by
\begin{equation}
\widetilde{H}=-\frac{1}{2}\frac{d^{2}}{dx^{2}}+V_{0}\left( e^{-2\alpha
x}-2e^{-\alpha x}\right) \text{, \ }x\in \mathbb{R}  \tag{3.1}
\end{equation}
where $V_{0}$ and $\alpha $ are nonnegative given parameters of the
oscillator. $H$ is an unbounded operator on the Hilbert space $L^{2}\left(
\mathbb{R}\right) $ and it is a selfadjoint operator with respect to its
form domain ($\left[ 4\right] $, Ch$.5)$. Note that $\min V\left( x\right)
=-V_{0}$, so that the discrete spectrum, if it exists, is contained in $%
\left[ -V_{0},0\right] .$

The Morse Hamiltonian is known to have a tridiagonal representation
representation in the orthogonal basis
\begin{equation}
\phi _{n}\left( x\right) :=\sqrt{\frac{n!\alpha }{\Gamma \left( n+2\gamma
+1\right) }}\left( \xi \right) ^{\gamma +\frac{1}{2}}\exp \left( -\frac{1}{2}%
\xi \right) L_{n}^{\left( 2\gamma \right) }\left( \xi \right) ,  \tag{3.3}
\end{equation}
$\xi :=\alpha ^{-1}\sqrt{8V_{0}}e^{-\alpha x}$ and $L_{n}^{\left( 2\gamma
\right) }\left( .\right) $ is the Laguerre polynomial. Here, $\gamma $ is
free scale parameter satisfying the restriction $2\gamma >-1$ but otherwise
arbitrary. The matrix representation of the Hamiltonian in this basis is
given explicitly as
\begin{equation}
\left\langle \phi _{n}\mid \widetilde{H}\mid \phi _{m}\right\rangle =%
\widetilde{b}_{n-1}\delta _{n,m+1}+\widetilde{a}_{n}\delta _{n,m}+\widetilde{%
b}_{n}\delta _{n,m-1}  \tag{3.4}
\end{equation}
where coefficients $\left( \widetilde{a}_{n}\right) $ and $\left( \widetilde{%
b}_{n}\right) $ are explicitly given by
\begin{equation}
\widetilde{a}_{n}=\frac{\alpha ^{2}}{2}\left( \left( n+\gamma +\frac{1}{2}%
-D\right) ^{2}+n\left( n+2\gamma \right) -D^{2}\right)  \tag{3.5}
\end{equation}
and
\begin{equation}
\widetilde{b}_{n}=-\frac{\alpha ^{2}}{2}\sqrt{\left( n+1\right) \left(
n+2\gamma +1\right) }\left( n+\gamma +\frac{1}{2}-D\right) \text{.}
\tag{3.6}
\end{equation}
We now expand the solutions $\left| \varphi _{E}\right\rangle $ of the
eigenvalue problem
\begin{equation}
\widetilde{H}\left| \varphi _{E}\right\rangle =E\left| \varphi
_{E}\right\rangle  \tag{3.7}
\end{equation}
in the basis $\left| \phi _{n}\right\rangle $ with real coefficients as
\begin{equation}
\left| \varphi _{E}\right\rangle =\sum\limits_{n=0}^{+\infty }\widetilde{%
\mathcal{C}}_{n}\left( E\right) \left| \phi _{n}\right\rangle .  \tag{3.8}
\end{equation}
Then, making use of $\left( 3.4\right) $ and the orthonormality of the basis
$\left| \phi _{n}\right\rangle $, one readily obtains the following
recurrence representation of the expansion coefficients
\begin{equation}
E\widetilde{\mathcal{C}}_{0}=a_{0}\widetilde{\mathcal{C}}_{0}+b_{0}%
\widetilde{\mathcal{C}}_{1}  \tag{3.9}
\end{equation}
\begin{equation}
E\widetilde{\mathcal{C}}_{n}\left( E\right) =b_{n-1}\widetilde{\mathcal{C}}%
_{n-1}\left( E\right) +a_{n}\widetilde{\mathcal{C}}_{n}\left( E\right) +b_{n}%
\widetilde{\mathcal{C}}_{n+1}\left( E\right) ,  \tag{3.10}
\end{equation}
for every integer $n=1,2,...$ . \ More explicitly, we have
\begin{equation*}
E\widetilde{\mathcal{C}}_{n}\left( E\right) =-\frac{\alpha ^{2}}{2}\sqrt{%
n\left( n+2\gamma \right) }\left( n-\frac{1}{2}+\gamma -D\right) \widetilde{%
\mathcal{C}}_{n-1}\left( E\right)
\end{equation*}
\begin{equation}
+\frac{\alpha ^{2}}{2}\left( \left( n+\gamma +\frac{1}{2}-D\right)
^{2}+n\left( n+2\gamma \right) -D^{2}\right) \widetilde{\mathcal{C}}%
_{n}\left( E\right)  \tag{3.11}
\end{equation}
\begin{equation*}
-\frac{\alpha ^{2}}{2}\sqrt{\left( n+1\right) \left( n+2\gamma +1\right) }%
\left( n+\gamma +\frac{1}{2}-D\right) \widetilde{\mathcal{C}}_{n+1}\left(
E\right) \text{.}
\end{equation*}

Before we proceed to find a solution to this recurrence relation, we show
that it contains information on two essential elements that we will need to
complete our analysis, namely, the ground state energy and the maximum
number of bound states that the Morse potential supports.

We recall that the three term recursion relation Eq.$\left( 3.10\right) $
can be cast in matrix form
\begin{equation}
\left(
\begin{array}{cccccccc}
\widetilde{a}_{0} & \widetilde{b}_{0} & 0 & 0 & 0 & 0 & 0 & 0 \\
\widetilde{b}_{0} & \widetilde{a}_{1} & \widetilde{b}_{1} & 0 & 0 & 0 & 0 & 0
\\
0 & \widetilde{b}_{1} & \widetilde{b}_{2} & \widetilde{b}_{2} & 0 & 0 & 0 & 0
\\
... & ... & ... & ... & ... & ... & ... & ... \\
0 & 0 & 0 & \widetilde{b}_{N-2} & \widetilde{a}_{N-1} & \widetilde{b}_{N-1}
& 0 & 0 \\
0 & 0 & 0 & 0 & \widetilde{b}_{N-1} & \widetilde{a}_{N} & \widetilde{b}_{N}
& 0 \\
0 & 0 & 0 & 0 & 0 & \widetilde{b}_{N} & \widetilde{b}_{N+1} & \widetilde{b}%
_{N+1} \\
... & ... & ... & ... & ... & ... & ... & ...
\end{array}
\right) \left(
\begin{array}{c}
\widetilde{\mathcal{C}}_{0}\left( E\right) \\
\widetilde{\mathcal{C}}_{1}\left( E\right) \\
\widetilde{\mathcal{C}}_{2}\left( E\right) \\
... \\
\widetilde{\mathcal{C}}_{N-1}\left( E\right) \\
\widetilde{\mathcal{C}}_{N}\left( E\right) \\
\widetilde{\mathcal{C}}_{N+1}\left( E\right) \\
...
\end{array}
\right) =E\left(
\begin{array}{c}
\widetilde{\mathcal{C}}_{0}\left( E\right) \\
\widetilde{\mathcal{C}}_{1}\left( E\right) \\
\widetilde{\mathcal{C}}_{2}\left( E\right) \\
... \\
\widetilde{\mathcal{C}}_{N-1}\left( E\right) \\
\widetilde{\mathcal{C}}_{N}\left( E\right) \\
\widetilde{\mathcal{C}}_{N+1}\left( E\right) \\
...
\end{array}
\right) \text{.}  \tag{3.12}
\end{equation}
Since the basis $\left( 3.3\right) $ is square integrable, any permissible
choice of parameters that makes the coefficient $b_{N-1}$ vanishes indicates
that the potential can support at least $N$ bound states whose coefficients $%
\widetilde{\mathcal{C}}_{n}$ for $n\geq N$ vanish identically and the
coefficients $\left\{ \widetilde{\mathcal{C}}_{n}\right\} _{n=0}^{N-1}$
satisfy the finite matrix equation
\begin{equation}
\left(
\begin{array}{ccccc}
\widetilde{a}_{0} & \widetilde{b}_{0} & 0 & 0 & 0 \\
\widetilde{b}_{0} & \widetilde{a}_{1} & \widetilde{b}_{1} & 0 & 0 \\
... & ... & ... & ... & ... \\
0 & 0 & \widetilde{b}_{N-3} & \widetilde{b}_{N-2} & \widetilde{b}_{N-2} \\
0 & 0 & 0 & \widetilde{b}_{N-2} & \widetilde{a}_{N-1}
\end{array}
\right) \left(
\begin{array}{c}
\widetilde{\mathcal{C}}_{0}\left( E\right) \\
\widetilde{\mathcal{C}}_{1}\left( E\right) \\
... \\
\widetilde{\mathcal{C}}_{N-2}\left( E\right) \\
\widetilde{\mathcal{C}}_{N-1}\left( E\right)
\end{array}
\right) =E\left(
\begin{array}{c}
\widetilde{\mathcal{C}}_{0}\left( E\right) \\
\widetilde{\mathcal{C}}_{1}\left( E\right) \\
... \\
\widetilde{\mathcal{C}}_{N-2}\left( E\right) \\
\widetilde{\mathcal{C}}_{N-1}\left( E\right)
\end{array}
\right) \text{.}  \tag{3.13}
\end{equation}
This is an eigenvalue equation yielding the energies of $N$ bound states $%
E_{\mu },\mu =0,1,...,N-1$ and corresponding eigenvectors $\left\{
\widetilde{\mathcal{C}}_{0}\left( E_{\mu }\right) ,\widetilde{\mathcal{C}}%
_{1}\left( E_{\mu }\right) ,...,\widetilde{\mathcal{C}}_{N-1}\left( E_{\mu
}\right) \right\} \mathcal{.}$ The total number of bound states is the
largest number $N$ for which we can make $b_{N-1}=0$, which means that $%
N+\gamma -\frac{1}{2}-D=0$. Such number is realized for the minimum value of
$\gamma $. Now, since $2\gamma >-1$, this immediately means that $N_{\max
}=\left\lfloor D+1\right\rfloor $, the integer part of the $D+1$, is the
number of bound states supported by the Morse potential.

On the other hand, as long as $D>0$ we can choose $\gamma $ to have the
specific value $\gamma =\gamma _{1}=D-1/2$. This makes $\widetilde{b}_{0}=0$%
. The above matrix equation reduces to the simple relation $\widetilde{a}_{0}%
\widetilde{\mathcal{C}}_{0}=E_{0}\widetilde{\mathcal{C}}_{0}.$ Hence, $%
E_{0}=-\frac{\alpha ^{2}}{2}D^{2}$ is the ground state energy. Incidintly,
the corresponding wave function from equation $\left( 3.8\right) $ is
\begin{equation}
\varphi _{E_{0}}\left( x\right) =\widetilde{\mathcal{C}}_{0}\phi ^{\left(
\gamma _{1}\right) }\left( x\right) =\sqrt{\frac{\alpha }{\Gamma \left(
2D\right) }}\xi ^{D}e^{-\frac{1}{2}\xi }\text{.}  \tag{3.14}
\end{equation}
By shifting the potential by $\frac{1}{2}\alpha ^{2}D^{2}$, the modified
Hamiltonian

\begin{equation}
H:=-\frac{1}{2}\frac{d^{2}}{dx^{2}}+V_{0}\left( e^{-2\alpha x}-2e^{-\alpha
x}\right) +\frac{1}{2}\alpha ^{2}D^{2}  \tag{3.15}
\end{equation}
will be semi-definite as is required by the formalism. \

The matrix representation of $H$ in the orthogonal basis
\begin{equation}
<\phi _{n}\mid H\mid \phi _{m}>=b_{n-1}\delta _{n,m+1}+a_{n}\delta
_{n,m}+b_{n}\delta _{n,m-1}  \tag{3.16}
\end{equation}
where $a_{n}=\widetilde{a}_{n}+\frac{1}{2}\alpha ^{2}D^{2}$ and $b_{n}=%
\widetilde{b}_{n}.$ We expand the solution $\mid \varphi _{\mathcal{E}}>$ of
the eigenvalue problem
\begin{equation}
H\mid \varphi _{\mathcal{E}}>=\mathcal{E}\mid \varphi _{\mathcal{E}}>
\tag{3.17}
\end{equation}
in the basis $\mid \phi _{n}>$ \ with real coefficients as
\begin{equation}
\mid \varphi _{\mathcal{E}}>=\sum\limits_{n=0}^{+\infty }\mathcal{C}%
_{n}\left( \mathcal{E}\right) \mid \phi _{n}>.  \tag{3.18}
\end{equation}
We note that $\mathcal{C}_{n}\left( \mathcal{E}\right) $ satisfies a
recursion relation similar to that satisfied by $\widetilde{\mathcal{C}}%
_{n}\left( E\right) $ except here $a_{n}$ replaces $\widetilde{a}_{n}$. It
turns out that this recursion relation is similar to that satisfied by the
continuous dual Hahn polynomials $S_{n}(t^{2};e,f,g)$ $\left( \left[ 5\right]
\text{, p.331}\right) $. Actually, we can show that
\begin{equation}
\mathcal{C}_{n}\left( \mathcal{E}\right) =\sqrt{\frac{\Gamma \left(
n+2\gamma +1\right) }{n!}}S_{n}\left( \lambda ^{2},-D,\gamma +\frac{1}{2}%
,\gamma +\frac{1}{2}\right)  \tag{3.19}
\end{equation}
with $\lambda ^{2}=2\alpha ^{-2}\mathcal{E}$ $-D^{2}$. The polynomials $%
\mathcal{P}_{n}\left( \mathcal{E}\right) $ defined by $\left( 2.16\right) $
are now obtained explicitly as
\begin{equation}
\mathcal{P}_{n}\left( \mathcal{E}\right) =\frac{\left( -1\right) ^{n}\left(
\gamma +\frac{1}{2}-D)\right) _{n}}{\sqrt{n!\left( 2\gamma +1\right) _{n}}}%
\text{ }_{3}\digamma _{2}\left(
\begin{array}{ccc}
-n & 1-D+i\lambda & 1-D-i\lambda \\
. & \gamma +\frac{1}{2}-D & \gamma +\frac{1}{2}-D
\end{array}
\mid 1\right)  \tag{3.20}
\end{equation}
in terms of a terminating $_{3}\digamma _{2}$-sum. The orthogonality
relations satisfied by the $\left\{ \mathcal{P}_{n}\right\} $ read
\begin{equation}
\delta _{j,k}=\int\limits_{\frac{\alpha ^{2}}{2}D^{2}}^{+\infty }\Omega
\left( \mathcal{E}\right) \mathcal{P}_{j}\left( \mathcal{E}\right) \overline{%
\mathcal{P}_{k}\left( \mathcal{E}\right) }d\mathcal{E}+\sum\limits_{m=0}^{%
\left\lfloor D\right\rfloor }\omega _{m}\mathcal{P}_{j}\left( \mathcal{E}%
_{m}\right) \overline{\mathcal{P}_{k}\left( \mathcal{E}_{m}\right) }
\tag{3.21}
\end{equation}
where
\begin{equation}
\Omega \left( \mathcal{E}\right) :=\left| \frac{\Gamma \left( -D+i\lambda
\left( \mathcal{E}\right) \right) \Gamma ^{2}\left( \gamma +\frac{1}{2}%
+i\lambda \left( \mathcal{E}\right) \right) }{2\pi \Gamma \left( 2i\lambda
\left( \mathcal{E}\right) \right) }\right| \frac{\left( \alpha ^{2}\lambda
\left( \mathcal{E}\right) \right) ^{-1}}{\Gamma ^{2}\left( \gamma +\frac{1}{2%
}-D\right) \Gamma \left( 2\gamma +1\right) }  \tag{3.22}
\end{equation}
\begin{equation}
\omega _{m}=\frac{\Gamma ^{2}\left( \gamma +\frac{1}{2}+D\right) }{\Gamma
\left( 2D\right) \Gamma \left( 2\gamma +1\right) }\frac{\left( -2D\right)
_{m}\left( -D+1\right) _{m}\left( \left( -D+\gamma +\frac{1}{2}\right)
_{m}\right) ^{2}\left( -1\right) ^{m}}{\left( -D\right) _{m}\left( \left(
-D-\gamma +\frac{1}{2}\right) _{m}\right) ^{2}m!}  \tag{3.23}
\end{equation}
and
\begin{equation}
\mathcal{E}_{m}=2^{-1}\alpha ^{2}\left( m\left( 2D-m\right) \right) ,\text{
\ \ \ }m=0,1,...,\left\lfloor D\right\rfloor  \tag{3.24}
\end{equation}
the energies of the bound states, as is shown in details in the appendix A.

\section{The supersymmetric partner Hamiltonian $H^{\left( +\right) }$}

Following the formalism we have summarized in section 2, we need \ first to
calculate the coefficients $\left( c_{n},d_{n}\right) $ given be their
expressions $\left( 2.18\right) $-$\left( 2.19\right) $ in terms of the
coefficients $b_{n}$\ and the values at zero of consecutive polynomials $%
\mathcal{P}_{n}.$ \ Here the $b_{n}$ are given by $\left( 3.6\right) $ and
we use $\left( 3.20\right) $ to calculate $\mathcal{P}_{j}\left( 0\right) $
as
\begin{equation}
\mathcal{P}_{j}\left( 0\right) =\left( -1\right) ^{j}\frac{\left( \gamma +%
\frac{1}{2}-D\right) _{j}}{\sqrt{j!\left( 2\gamma +1\right) _{j}}}\text{.}
\tag{4.1}
\end{equation}
This gives the following expressions for the coefficients $\left(
c_{n},d_{n}\right) :$
\begin{equation}
c_{n}=\frac{\alpha }{\sqrt{2}}\left( n+\gamma +\frac{1}{2}-D\right)
\tag{4.2}
\end{equation}
\begin{equation}
d_{n+1}=-\frac{\alpha }{\sqrt{2}}\sqrt{\left( n+1\right) \left( n+1+2\gamma
\right) }\text{.}  \tag{4.3}
\end{equation}
Next, we have shown that the partner Hamiltonian $H^{\left( +\right)
}=AA^{\dagger }$\ has a tridiagonal representation with respect to the same
basis $\mid \phi _{n}>$ with the form $\left( 2.7\right) \ $where the
coefficients $a_{n}^{\left( +\right) },b_{n}^{\left( +\right) }$ have been
expressed in terms coefficients $\left( c_{n},d_{n}\right) $ through the
relations $\left( 2.8\right) $-$\left( 2.9\right) $. Therefore, using
expressions $\left( 4.2\right) $-$\left( 4.3\right) ,$we obtain that\textbf{%
\ }
\begin{equation}
a_{n}^{\left( +\right) }=\frac{\alpha ^{2}}{2}\left( \left( n+1\right)
\left( n+2\gamma +1\right) +\left( n+\gamma +\frac{1}{2}-D\right)
^{2}\right)   \tag{4.4}
\end{equation}
\begin{equation}
\mathit{\ }b_{n}^{\left( +\right) }=-\frac{\alpha ^{2}}{2}\sqrt{\left(
n+1\right) \left( n+2\gamma +1\right) }\left( n+\gamma +\frac{3}{2}-D\right)
\tag{4.5}
\end{equation}
for every $n=0,1,2,...$ . Note that these expressions are identical to those
of $\left( a_{n},b_{n}\right) $ except for the replacement of the parameter $%
-D$ by $-D+1$.

This last property is powerful. It is equivalent to the shape-invariance
property $\left[ 6\right] $ that enables one to deduce the properties of the
supersymmetric partner Hamiltonian $H^{\left( +\right) }$ from those of $H$.
In our case, we can immediately deduce the following.

$\left( i\right) $ The energy spectrum $\left\{ \mathcal{E}_{m}^{\left(
+\right) }\right\} $ can be obtained from the spectrum $\left\{ \mathcal{E}%
_{m}\right\} $ by the replacement $-D\rightarrow -D+1.$ More specifically,
\begin{equation}
\mathcal{E}_{m}^{\left( +\right) }=-\left( \left( D-1\right) -m\right)
^{2}=-\left( D-\left( m+1\right) \right) ^{2}=\mathcal{E}_{m+1}  \tag{4.6}
\end{equation}

$\left( ii\right) $ $\mathcal{P}_{n}^{\left( +\right) }$ can be deduced \
from the solution $\left( 3.13\right) $ for $\mathcal{P}_{n}$ as
\begin{equation}
\mathcal{P}_{n}^{\left( +\right) }\left( \mathcal{E}\right) =\frac{\left(
-1\right) ^{n}\left( \gamma +\frac{3}{2}-D)\right) _{n}}{\sqrt{n!\left(
2\gamma +1\right) _{n}}}\text{ }_{3}F_{2}\left(
\begin{array}{ccc}
-n & 1-D+i\lambda & 1-D-i\lambda \\
. & \gamma +\frac{3}{2}-D & \gamma +\frac{3}{2}-D
\end{array}
\mid 1\right)  \tag{4.7}
\end{equation}
We could have proceeded of finding the solution for $\mathcal{P}_{n}^{\left(
+\right) }$ from equation $\left( 2.21\right) $. In appendix B, we do jsut
that and show that we get identical results.

$\left( iii\right) $ The orthogonality relations satisfied by the $\mathcal{P%
}_{n}^{\left( +\right) }$ are now given by

\begin{equation}
\delta _{j,k}=\int\limits_{\frac{\alpha ^{2}}{2}\left( D-1\right)
^{2}}^{+\infty }\Omega ^{\left( +\right) }\left( \mathcal{E}\right) \mathcal{%
P}_{j}^{\left( +\right) }\left( \mathcal{E}\right) \overline{\mathcal{P}%
_{k}^{\left( +\right) }\left( \mathcal{E}\right) }d\mathcal{E}%
+\sum\limits_{m=0}^{\left\lfloor D-1\right\rfloor }\omega _{m}^{\left(
+\right) }\mathcal{P}_{j}^{\left( +\right) }\left( \mathcal{E}_{m}\right)
\overline{\mathcal{P}_{k}^{\left( +\right) }\left( \mathcal{E}_{m}\right) }
\tag{4.8}
\end{equation}
where
\begin{equation}
\Omega ^{\left( +\right) }\left( \mathcal{E}\right) =\left| \frac{\Gamma
\left( 1-D+i\lambda \left( \mathcal{E}\right) \right) \Gamma ^{2}\left(
\gamma +\frac{1}{2}+i\lambda \left( \mathcal{E}\right) \right) }{2\pi \Gamma
\left( 2i\lambda \left( \mathcal{E}\right) \right) }\right| \frac{\left(
\alpha ^{2}\lambda \left( \mathcal{E}\right) \right) ^{-1}}{\Gamma
^{2}\left( \gamma +\frac{3}{2}-D\right) \Gamma \left( 2\gamma +1\right) }
\tag{4.9}
\end{equation}
and
\begin{equation}
\omega _{m}^{\left( +\right) }=\frac{\Gamma ^{2}\left( \gamma +\frac{3}{2}%
+D\right) }{\Gamma \left( 2D-2\right) \Gamma \left( 2\gamma +1\right) }\frac{%
\left( -2D+2\right) _{m}\left( -D+2\right) _{m}\left( \left( -D+\gamma +%
\frac{3}{2}\right) _{m}\right) ^{2}}{\left( -1\right) ^{m}m!\left(
-D+1\right) _{m}\left( \left( -D-\gamma +\frac{3}{2}\right) _{m}\right) ^{2}}
\tag{4.10}
\end{equation}

\section{Conclusion}

In $\left[ 1\right] $ we have been concerned with the supersymmetry of a
positive semi-definite Hamiltonian given by $H=A^{\dagger }A$ in terms of a
forward-shift operator $A$ and its adjoint $A^{\dagger }$, which was assumed
to have a tridiagonal matrix representation in a chosen basis. We have
proved that the supersymmetirc partner Hamiltonian $H^{\left( +\right) }=A$ $%
A^{\dagger }$ also have a tridiagonal matrix representation in the same
basis and that the polynomials associated with the eigenstates expansion of $%
H^{\left( +\right) }$ are precisely the kernel polynomials of those
associated with $H$. These results have been illustrated to the free
particle whose spectrum is purely continuous and the harmonic oscilator
whose spectrum is purley discrete. As an example for which the method works,
we here have been dealing with the Morse oscillator which may have a finite
discrete spectrum in addition to a continuous one and moreover there exists
a basis that renders it tridiagonal. To conclude, we now can confirm that
the results obtainted here with the previous ones complete the treatment of
tridiagonal Hamiltonians with pure continuous, or a pure discrete one, or a
spectrum of mixed discrete and continous spectra. So that our analysis could
be applied to a large family of Hamiltonians for which there exist square
integrable bases supporting their infinite tridiagonal matrix representation
(see $\left[ 7\right] $ and references therein).

\newpage

\section{Appendix A}

In order to discuss the orthogonality relations of the obtained polynomials $%
\left\{ \mathcal{P}_{n}\left( \mathcal{E}\right) \right\} $ in $\left(
3.21\right) ,$ we recall the following properties of continuous dual Hahn
polynomials $\left[ 5\right] $. These polynomials are defined by

\begin{equation}
S_{n}\left( y,a,b,c\right) =\left( -1\right) ^{n}\sqrt{\frac{\left(
a+b\right) _{n}\left( a+c\right) _{n}}{n!\left(b+c\right) _{n}}}\text{ }%
_{3}F_{2}\left(
\begin{array}{ccc}
-n & a+it & a-it \\
. & a+b & a+c
\end{array}
\mid 1\right) ,\text{ \ }  \tag{A1}
\end{equation}
where $t^{2}=y$ and $n=0,1,2,...$, and are symmetric in $a,b$ and $c.$
Assume that $a$ is the smallest of the real parameters $a,b$ and $c.$ \ Let $%
d\mu \left( \cdot ,a,b,c\right) $ to the measure defined by
\begin{equation}
\int\limits_{\mathbb{R}}f\left( y\right) d\mu \left( y\right) =\frac{1}{2\pi
}\int\limits_{0}^{+\infty }\left| \frac{\Gamma \left( a+it\right) \Gamma
\left( b+it\right) \Gamma \left( c+it\right) }{\Gamma \left( 2it\right) }%
\right| \frac{f\left( t^{2}\right) }{\Gamma \left( a+b\right) \Gamma \left(
a+c\right) \Gamma \left( b+c\right) }dt  \tag{A2}
\end{equation}
\begin{equation*}
+\frac{\Gamma \left( b-a\right) \Gamma \left( c-a\right) }{\Gamma \left(
-2a\right) \Gamma \left( b+c\right) }\sum\limits_{l=0}^{K}\frac{\left(
2a\right) _{l}\left( a+1\right) _{l}\left( a+b\right) _{l}\left( a+c\right)
_{l}}{\left( a\right) _{l}\left( a-b+1\right) _{l}\left( a-c+1\right) _{l}}%
\frac{\left( -1\right) ^{l}}{l!}f\left( -\left( a+l\right) ^{2}\right) ,
\end{equation*}
where $K$ is the largest non-negative integer such that $a+K<0$. The measure
$d\mu \left( \cdot ,a,b,c\right) $ is absolutely continuous if $a\geq 0.$
The measure is positive under the conditions $a+b>0$, $a+c>0$ and $b+c>0$.
Then the polynomials $S_{n}\left( y,a,b,c\right) $ are orthonormal with
respect to the measure $d\mu \left( y,a,b,c\right) $.

Accoording to Eq.$\left( \text{A2}\right) $ the \textit{continuous}
orthogonality reads in our case
\begin{equation}
\frac{1}{2\pi }\int\limits_{0}^{+\infty }\left| \frac{\Gamma \left(
a+i\lambda \right) \Gamma \left( b+i\lambda \right) \Gamma \left( c+i\lambda
\right) }{\Gamma \left( 2i\lambda \right) }\right| \frac{\mathcal{P}%
_{j}\left( \frac{\alpha ^{2}}{2}\left( D^{2}+\lambda ^{2}\right) \right)
\overline{\mathcal{P}_{k}\left( \frac{\alpha ^{2}}{2}\left( D^{2}+\lambda
^{2}\right) \right) }}{\Gamma \left( a+b\right) \Gamma \left( a+c\right)
\Gamma \left( b+c\right) }d\lambda \text{.}  \tag{A3}
\end{equation}
This integration can be rewritten is terms of the energy variable $\mathcal{E%
}\equiv \mathcal{E}\left( \lambda \right) =2^{-1}\alpha ^{2}\left( \lambda
^{2}+D^{2}\right) $ where $\lambda \geq 0$ (can be expressed as $\lambda
\equiv \lambda \left( \mathcal{E}\right) )$ and parameters $a=-D$ and $%
b=c=\gamma +\frac{1}{2}$ as
\begin{equation}
\int\limits_{\frac{\alpha ^{2}}{2}D^{2}}^{+\infty }\left| \frac{\Gamma
\left( -D+i\lambda \left( \mathcal{E}\right) \right) \Gamma ^{2}\left(
\gamma +\frac{1}{2}+i\lambda \left( \mathcal{E}\right) \right) }{\Gamma
\left( 2i\lambda \left( \mathcal{E}\right) \right) }\right| \frac{\mathcal{P}%
_{j}\left( \mathcal{E}\right) \overline{\mathcal{P}_{k}\left( \mathcal{E}%
\right) }\left( \alpha ^{2}\lambda \left( \mathcal{E}\right) \right) ^{-1}}{%
2\pi \Gamma ^{2}\left( \gamma +\frac{1}{2}-D\right) \Gamma \left( 2\gamma
+1\right) }dE  \tag{A4}
\end{equation}
Therefore, we consider the density function
\begin{equation}
\Omega \left( E\right) :=\left| \frac{\Gamma \left( -D+i\lambda \left(
\mathcal{E}\right) \right) \Gamma ^{2}\left( \gamma +\frac{1}{2}+i\lambda
\left( \mathcal{E}\right) \right) }{2\pi \Gamma \left( 2i\lambda \left(
\mathcal{E}\right) \right) }\right| \frac{\left( \alpha ^{2}\lambda \left(
\mathcal{E}\right) \right) ^{-1}}{\Gamma ^{2}\left( \gamma +\frac{1}{2}%
-D\right) \Gamma \left( 2\gamma +1\right) }.  \tag{A5}
\end{equation}
On the other hand, according to $\left( \text{A2}\right) $ the discrete\
orthogonality reads in our case

\begin{equation}
\sum\limits_{l=0}^{K}\omega _{l}P_{j}\left( \frac{\alpha ^{2}}{2}\left(
D^{2}+(i\left( a+l\right) )^{2}\right) \right) \overline{P_{k}\left( \frac{%
\alpha ^{2}}{2}\left( D^{2}+(i\left( a+k\right) )^{2}\right) \right) }
\tag{A6}
\end{equation}
where
\begin{equation}
\omega _{l}:=\frac{\Gamma \left( b-a\right) \Gamma \left( c-a\right) }{%
\Gamma \left( -2a\right) \Gamma \left( b+c\right) }\frac{\left( 2a\right)
_{l}\left( a+1\right) _{l}\left( a+b\right) _{l}\left( a+c\right) _{l}\left(
-1\right) ^{l}}{\left( a\right) _{l}\left( a-b+1\right) _{l}\left(
a-c+1\right) _{l}l!}  \tag{A7}
\end{equation}
with parameters $a=-D,b=c=\gamma +\frac{1}{2}$ and $K=\left\lfloor
D\right\rfloor .$ By changing the summation index $l=m$ and recalling that
the discrete spectrum of $H$ is given by the eigenvalues $\mathcal{E}%
_{m}=2^{-1}\alpha ^{2}\left( m\left( 2D-m\right) \right) $ for $%
m=0,1,...,\left\lfloor D\right\rfloor $ then $\left( \text{A6}\right) $
takes the form
\begin{equation}
\sum\limits_{m=0}^{\left[ D\right] }\omega _{m}\mathcal{P}_{j}\left(
\mathcal{E}_{m}\right) \overline{\mathcal{P}_{k}\left( \mathcal{E}%
_{m}\right) }.  \tag{A8}
\end{equation}
where
\begin{equation}
\omega _{m}=\frac{\Gamma ^{2}\left( \gamma +\frac{1}{2}+D\right) }{\Gamma
\left( 2D\right) \Gamma \left( 2\gamma +1\right) }\frac{\left( -2D\right)
_{m}\left( -D+1\right) _{m}\left( \left( -D+\gamma +\frac{1}{2}\right)
_{m}\right) ^{2}}{\left( -1\right) ^{m}m!\left( -D\right) _{m}\left( \left(
-D-\gamma +\frac{1}{2}\right) _{m}\right) ^{2}}\text{.}  \tag{A9}
\end{equation}

\section{Appendix B}

We now proceed to find the polynomial $\mathcal{P}_{n}^{\left( +\right)
}\left( \mathcal{E}\right) $ defined in $\left( 2.22\right) $ in terms of
the kernel polynomial
\begin{equation}
\mathcal{K}_{n}\left( \mathcal{E},0\right) =\sum\limits_{j=0}^{n}\mathcal{P}%
_{j}\left( \mathcal{E}\right) \mathcal{P}_{j}\left( 0\right)   \tag{B1}
\end{equation}
For this, we make use of $\left( 4.1\right) $ therefore $\left( \text{B1}%
\right) $ takes the form

\begin{equation}
\mathcal{K}_{n}\left( \mathcal{E},0\right) =\sum\limits_{j=0}^{n}\frac{%
\left( \left( \gamma +\frac{1}{2}-D\right) _{j}\right) ^{2}}{j!\left(
2\gamma +1\right) _{j}}\text{ }_{3}\digamma _{2}\left(
\begin{array}{ccc}
-j & -D+i\lambda \left( \mathcal{E}\right) & -D-i\lambda \left( \mathcal{E}%
\right) \\
. & \gamma +\frac{1}{2}-D & \gamma +\frac{1}{2}-D
\end{array}
\mid 1\right)  \tag{B2}
\end{equation}
\begin{equation}
=\sum\limits_{j=0}^{n}\frac{\Gamma ^{2}\left( \gamma +\frac{1}{2}-D+j\right)
\Gamma \left( 2\gamma +1\right) }{\Gamma ^{2}\left( \gamma +\frac{1}{2}%
-D\right) \Gamma \left( 2\gamma +1+j\right) }\frac{1}{j!}\text{ }%
_{3}\digamma _{2}\left(
\begin{array}{ccc}
-j & -D+i\lambda \left( \mathcal{E}\right) & -D-i\lambda \left( \mathcal{E}%
\right) \\
. & \gamma +\frac{1}{2}-D & \gamma +\frac{1}{2}-D
\end{array}
\mid 1\right)  \tag{B3}
\end{equation}
Now, make use of the following Thomae transformation $\left( \left[ 8\right]
\text{, p.186}\right) :$%
\begin{equation}
_{3}\digamma _{2}\left(
\begin{array}{ccc}
a & b & c \\
. & d & e
\end{array}
\mid 1\right) =\frac{\Gamma \left( e\right) \Gamma \left( d+e-a-b-c\right) }{%
\Gamma \left( e-a\right) \Gamma \left( d+e-b-c\right) }\text{ }_{3}\digamma
_{2}\left(
\begin{array}{ccc}
a & d-b & d-c \\
. & d & d+e-b-c
\end{array}
\mid 1\right)  \tag{B4}
\end{equation}
for the parameters $a=-j,b=-D+i\lambda \left( \mathcal{E}\right) ,c=-D-ix$
and $d=e=\gamma +\frac{1}{2}-D$. Here $\lambda \equiv \lambda \left(
\mathcal{E}\right) =\sqrt{\frac{2}{\alpha ^{2}}\mathcal{E}-D^{2}}.$ The $%
_{3}\digamma _{2}-$sum in $\left( \text{B3}\right) $ transforms to
\begin{equation}
\frac{\Gamma \left( \gamma +\frac{1}{2}-D\right) \Gamma \left( 2\gamma
+1+j\right) }{\Gamma \left( \gamma +\frac{1}{2}-D+j\right) \Gamma \left(
2\gamma +1\right) }\text{ }_{3}\digamma _{2}\left(
\begin{array}{ccc}
-j & \gamma +\frac{1}{2}-i\lambda \left( \mathcal{E}\right) & \gamma +\frac{1%
}{2}+i\lambda \left( \mathcal{E}\right) \\
. & \gamma +\frac{1}{2}-D & 2\gamma +1
\end{array}
\mid 1\right)  \tag{B5}
\end{equation}
Summarizing the above calculations, we arrive at
\begin{equation}
\mathcal{K}_{n}\left( \mathcal{E},0\right) =\sum\limits_{j=0}^{n}\frac{%
\left( \gamma +\frac{1}{2}-D\right) _{j}}{j!}\text{ }_{3}\digamma _{2}\left(
\begin{array}{ccc}
-j & \gamma +\frac{1}{2}-i\lambda \left( \mathcal{E}\right) & \gamma +\frac{1%
}{2}+i\lambda \left( \mathcal{E}\right) \\
. & \gamma +\frac{1}{2}-D & 2\gamma +1
\end{array}
\mid 1\right)  \tag{B6}
\end{equation}
Next, we make appeal to the summation formula $\left( \left[ 9\right] \text{%
, p.388, Eq.6}\right) $:
\begin{equation}
\sum\limits_{j=0}^{n}\frac{\left( \sigma \right) _{j}}{j!}\text{ }%
_{p+1}\digamma _{q}\left(
\begin{array}{cc}
-j & \left( a_{p}\right) \\
. & \left( b_{p}\right)
\end{array}
\mid t\right) =\frac{\left( \sigma +1\right) _{n}}{n!}\text{ }_{p+2}\digamma
_{q+1}\left(
\begin{array}{ccc}
-n & \sigma & \left( a_{p}\right) \\
. & \sigma +1 & \left( b_{p}\right)
\end{array}
\mid t\right)  \tag{B7}
\end{equation}
for $\sigma =\gamma +\frac{1}{2}-D,$ $p=2,q=2$ and $t=1,$ to get that

\begin{equation}
\mathcal{K}_{n}\left( \mathcal{E},0\right) =\frac{\left( \gamma +\frac{3}{2}%
-D\right) _{n}}{n!}\text{ }_{4}\digamma _{3}\left(
\begin{array}{cccc}
-n & \gamma +\frac{1}{2}-D & \gamma +\frac{1}{2}-i\lambda \left( \mathcal{E}%
\right) & \gamma +\frac{1}{2}+i\lambda \left( \mathcal{E}\right) \\
. & \gamma +\frac{3}{2}-D & \gamma +\frac{1}{2}-D & 2\gamma +1
\end{array}
\mid 1\right) .  \tag{B8}
\end{equation}
Because of the particular relation of the parameters of the $_{4}\digamma
_{3}$-sum, \ Eq.$\left( \text{B8}\right) $ reduces to
\begin{equation}
\mathcal{K}_{n}\left( \mathcal{E},0\right) =\frac{\left( \gamma +\frac{3}{2}%
-D\right) _{n}}{n!}\text{ }_{3}F_{2}\left(
\begin{array}{ccc}
-n & \gamma +\frac{1}{2}-i\lambda \left( \mathcal{E}\right) & \gamma +\frac{1%
}{2}+i\lambda \left( \mathcal{E}\right) \\
. & \gamma +\frac{3}{2}-D & 2\gamma +1
\end{array}
\mid 1\right)  \tag{B9}
\end{equation}
Once again, we can use the transformation $\left( \text{B4}\right) $ to
rewrite the $_{3}\digamma _{2}$-sum in $\left( \text{B9}\right) $ as
\begin{equation}
\frac{\Gamma \left( 2\gamma +1\right) \Gamma \left( \gamma +\frac{3}{2}%
-D+n\right) }{\Gamma \left( 2\gamma +1+n\right) \Gamma \left( -3\gamma -D-%
\frac{1}{2}\right) }\text{ }_{3}F_{2}\left(
\begin{array}{ccc}
-n & 1-D+i\lambda \left( \mathcal{E}\right) & 1-D-i\lambda \left( \mathcal{E}%
\right) \\
. & \gamma +\frac{3}{2}-D & \gamma +\frac{3}{2}-D
\end{array}
\mid 1\right)  \tag{B10}
\end{equation}
Therefore $\left( \text{B9}\right) $ becomes
\begin{equation}
\mathcal{K}_{n}\left( \mathcal{E},0\right) =\frac{\left( \gamma +\frac{1}{2}%
+(-D+1)\right) _{n}}{n!\left( 2\gamma +1\right) _{n}}\frac{\Gamma \left(
\gamma +\frac{3}{2}-D+n\right) }{\Gamma \left( \gamma +\frac{3}{2}-D\right) }
\tag{B11}
\end{equation}
\begin{equation*}
\times _{3}F_{2}\left(
\begin{array}{ccc}
-n & (-D+1)+i\lambda \left( \mathcal{E}\right) & (-D+1)-i\lambda \left(
\mathcal{E}\right) \\
. & \gamma +\frac{1}{2}+(-D+1) & \gamma +\frac{1}{2}+(-D+1)
\end{array}
\mid 1\right)
\end{equation*}
Since
\begin{equation}
\sqrt{\frac{b_{0}\mathcal{P}_{1}\left( 0\right) }{b_{n}\mathcal{P}_{n}\left(
0\right) \mathcal{P}_{n+1}\left( 0\right) }}=\frac{\left( -1\right)
^{n}\left( \gamma +\frac{1}{2}-D\right) \sqrt{n!\left( 2\gamma +1\right) _{n}%
}}{\left( \gamma +\frac{1}{2}-D\right) _{n}\left( \gamma +\frac{1}{2}%
-D+n\right) }  \tag{B12}
\end{equation}
the above results can be substitued in equation $\left( 2.22\right) $ to
give
\begin{equation}
\mathcal{P}_{n}^{\left( +\right) }\left( \mathcal{E}\right) =\frac{\left(
-1\right) ^{n}\left( \gamma +\frac{3}{2}-D)\right) _{n}}{\sqrt{n!\left(
2\gamma +1\right) _{n}}}._{3}F_{2}\left(
\begin{array}{ccc}
-n & 1-D+i\lambda \left( \mathcal{E}\right) & 1-D-i\lambda \left( \mathcal{E}%
\right) \\
. & \gamma +\frac{3}{2}-D & \gamma +\frac{3}{2}-D
\end{array}
\mid 1\right)  \tag{B13}
\end{equation}
which is an expected result confirming the same correspondence $%
-D\rightarrow -D+1$ or $D\rightarrow \left( D-1\right) $ noted earlier.

\newpage

\textbf{References}

$\left[ 1\right] $\ Yamani H A and Mouayn Z 2014 Supersymmetry of
tridiagonal Hamiltonians\ \textit{J. Phys. A: Math. Theor}. \textbf{47} doi:
10.1088/1751-8113/47/26/265203

$\left[ 2\right] $\ Szeg\"{o} G 1939 Orthogonal Polynomials, Colloquium
Publications vol 23 (Providence, RI: American Mathematical Society)

$\left[ 3\right] $\ Chihara S T 1978 An Introduction to Orthogonal
Polynomials (London: Gordon and Breach)

$\left[ 4\right] $\ Schechter M 1981 Operator Methods in Quantum Mechanics
(New York: North-Holland)

$\left[ 5\right] $\ Andrews G E, Askey R and Roy R 1999 Special Functions,
Encycl. Math. Appl. 71, Cambridge Univ. Press

$\left[ 6\right] $\ Gendenshtein L E 1983 Derivation of exact spectra of the
Schr\"{o}dinger equation by means of supersymmetry \textit{JETP Lett.}
\textbf{38} (6) 356

$\left[ 7\right] $\ Alhaidari A D 2005 An extended class of $L^{2}$-series
solutions of the wave equation, \textit{Ann. Phys.}\textbf{\ 317} 152

$\left[ 8\right] $\ Krattenthaler C and Rivoal T 2006 How can we escape
Thomae's relations ? \textit{J. Math. Soc. Japan} \textbf{58} (1) 183

$\left[ 9\right] $\ Prudnikov A P, Brychkov Yu A and Marichev O I 1990
Integrals and Series Vol.3 (Amsterdam: Gordon and Breach)

{\footnotesize \bigskip }

\end{document}